\documentclass{article} 
\usepackage{iclr2027_conference,times}

\usepackage{amsmath,amsfonts,bm}

\def\eqref#1{equation~\ref{#1}}

\def\1{\bm{1}}

\DeclareMathAlphabet{\mathsfit}{\encodingdefault}{\sfdefault}{m}{sl}
\SetMathAlphabet{\mathsfit}{bold}{\encodingdefault}{\sfdefault}{bx}{n}

\usepackage{hyperref}
\usepackage{url}
\usepackage{graphicx} 
\usepackage{booktabs}
\usepackage{multirow}
\usepackage{amssymb}
\usepackage{fancyvrb}
\usepackage{enumitem}
\newcommand{\best}[1]{\textbf{#1}}
\newcommand{\second}[1]{\underline{#1}}

\usepackage[T1]{fontenc}
\usepackage{textcomp}

\usepackage{xcolor}
\usepackage{tcolorbox}
\tcbuselibrary{breakable}

\usepackage{upquote}
\usepackage{listings}
\usepackage{titletoc}
\usepackage{placeins}

\lstdefinestyle{promptstyle}{
    basicstyle=\fontfamily{lmtt}\selectfont\scriptsize,
    columns=fullflexible,
    keepspaces=true,
    breaklines=true,
    breakatwhitespace=false,
    breakindent=0pt,
    breakautoindent=false,
    postbreak=\mbox{},
    upquote=true,
    aboveskip=2pt,
    belowskip=0pt,
    xleftmargin=0pt,
}

\definecolor{cspromptc}{HTML}{2563EB}

\newtcolorbox{promptcard}[1]{
    breakable,
    colback=white,
    colframe=cspromptc,
    boxrule=0.9pt,
    arc=2.5pt,
    left=6pt,
    right=6pt,
    top=4pt,
    bottom=4pt,
    fonttitle=\bfseries\small,
    coltitle=white,
    colbacktitle=cspromptc,
    title={#1}
}

\newcounter{promptbox}[section]
\renewcommand{\thepromptbox}{\thesection.\arabic{promptbox}}

\newcommand{\promptbox}[3]{%
    \refstepcounter{promptbox}%
    \label{#3}%
    \begin{promptcard}{Prompt~\thepromptbox: #1}
        \lstinputlisting[style=promptstyle]{#2}
    \end{promptcard}
}

\definecolor{csfilec}{HTML}{52796F}

\newtcolorbox{filecard}[1]{%
    breakable,
    colback=white,
    colframe=csfilec,
    boxrule=0.9pt,
    arc=2.5pt,
    left=6pt,
    right=6pt,
    top=4pt,
    bottom=4pt,
    fonttitle=\bfseries\small,
    coltitle=white,
    colbacktitle=csfilec,
    title={#1}
}

\newcommand{\filebox}[2]{%
    \begin{filecard}{#1}
        \lstinputlisting[style=promptstyle]{#2}
    \end{filecard}
}

\title{\textsc{Harde}: Optimizing Agent Harnesses for Runtime Risk Detection and Execution Control}

\author{
Zhuo Liu\textsuperscript{1},
Moxin Li\textsuperscript{2},
Zhixin Ma\textsuperscript{3},
Wentao Shi\textsuperscript{1},
Wenjie Wang\textsuperscript{1},
Fuli Feng\textsuperscript{1}
\\
\textsuperscript{1}University of Science and Technology of China \\
\textsuperscript{2}National University of Singapore
\quad
\textsuperscript{3}Singapore Management University \\
}

\iclrfinalcopy 
\begin{document}

\maketitle

\fancyhead{}
\renewcommand{\headrulewidth}{0pt}

\begin{abstract}

Large language model (LLM) agents are vulnerable to safety risks such as injected malicious instructions or misleading information, motivating \emph{runtime defenses} that prevent unsafe action in execution across diverse risks while preserving benign-task utility.
Existing system-level defenses either focus on risk detection rather than timely prevention or rely on predefined rules with limited flexibility across diverse risks. 
We propose a risk-aware harness that integrates LLM-based monitoring for flexible risk detection and structures monitor-guided execution around three core modules—trigger, monitor, and feedback—enabling targeted safety interventions while limiting disruption to benign task execution.
To adapt the harness to different risks and deployment settings, we introduce \textsc{Harde}, a two-stage harness optimization framework that first performs isolated probing of each module to derive an optimization guide, then uses this guide to iteratively optimize the harness based on safety and utility feedback.
Experiments across three attack benchmarks show that \textsc{Harde} improves runtime safety while preserving utility, outperforming manually designed harnesses and naive optimization baselines.
Our analysis shows that effective runtime defense benefits from complementary safety mechanisms, attack-aware harness optimization, and harness designs matched to monitor capabilities. Our code is available at \href{https://github.com/Liuz233/HARDE}{\underline{this repository}}.

\end{abstract}

\section{Introduction}
\label{sec:introduction}

Large language model (LLM) agents are autonomous systems that interact with external environments via observation, reasoning and planning, and taking actions~\citep{yao2022react,liu2024agentbench,schick2023toolformer}. However, such interactions also introduce safety risks. For example, agents can be manipulated through untrusted environmental signals containing misleading information~\citep{liao2025eia,zhang2024agent} or malicious instructions~\citep{greshake2023not,debenedetti2024agentdojo}, potentially leading to severe real-world consequences~\citep{andriushchenko2025agentharm}. This motivates the development of \emph{defense mechanisms}~\citep{lin2026safeharness,mou2026toolsafe} that detect and prevent unsafe agent behaviors while preserving utility on benign tasks.

Existing defense approaches for LLM agents can be broadly categorized into \emph{model-level alignment} and \emph{system-level defenses}. Model-level approaches align the underlying model through supervised fine-tuning or reinforcement learning to reduce unsafe  actions~\citep{zhang2025agentalign,sha2025agent,liu2026safeagent}, but incur substantial computational costs for initial training and continual updates. 
System-level defenses monitor the agent execution loop and enforce runtime constraints, 
without retraining the agent model, and are the focus of this work.
Existing approaches mainly follow two directions: 
LLM-based monitors that classify agent execution trajectories as safe or unsafe using separate lightweight models~\citep{NEURIPS2025_468a018d,kale2025reliable,guan2025monitoring}, and safety harnesses that enforce execution-time constraints with mechanisms such as
rule-based filtering~\citep{wang2025agentspec},
sandboxing~\citep{wu2025isolategpt} or privilege separation~\citep{lin2026safeharness},
tool-access control~\citep{debenedetti2025defeating},
and runtime intervention~\citep{mou2026toolsafe}.

However, current system-level defenses still struggle to provide effective \textbf{\emph{runtime defenses}} that require timely intervention before executing unsafe actions and flexibility across diverse risks. 
LLM-based monitors can identify diverse risks in agent trajectories but focus on risk detection rather than how detection should be integrated with execution-time intervention~\citep{NEURIPS2025_468a018d}. 
Safety harnesses can enforce runtime interventions, but typically rely on predefined rules for risk detection and control, limiting their flexibility across diverse risks~\citep{lin2026safeharness}. Therefore, we argue that effective runtime defense requires a risk-aware agent execution harness that leverages flexible risk detection to control execution.
Such a harness can selectively invoke LLM-based monitors to detect potential risks and control the execution through targeted interventions, such as blocking actions or editing execution contexts, while preserving benign task completion. 
Achieving both safety and benign-task utility across different risks and deployment settings requires adapting the harness’s monitoring and intervention strategies, motivating their automated optimization~\citep{lee2026meta}.


Realizing such a risk-aware harness poses two key challenges. First, balancing runtime safety and benign-task utility requires coordinating when to invoke monitoring, how to detect risks, and how to intervene in agent execution to enable targeted safety interventions while limiting disruption to benign task execution.
These lead to a large harness design space. We therefore structure monitor-guided execution around three core modules: \textbf{\textit{trigger module}}, which determines when monitoring is invoked for potential risks; \textbf{\textit{monitor module}}, which detects risks in proposed actions based on the execution context; and \textbf{\textit{feedback module}}, which determines how to sanitize agent actions or execution contexts. 
Second, effectively optimizing these coupled modules is challenging because the effects of individual modules on runtime safety and task completion remain ambiguous. We therefore perform probing of each module prior to optimization to isolate its effects, identify recurring failure modes, and guide subsequent harness optimization.

\begin{figure}[t]
    \centering
    \includegraphics[width=\textwidth]{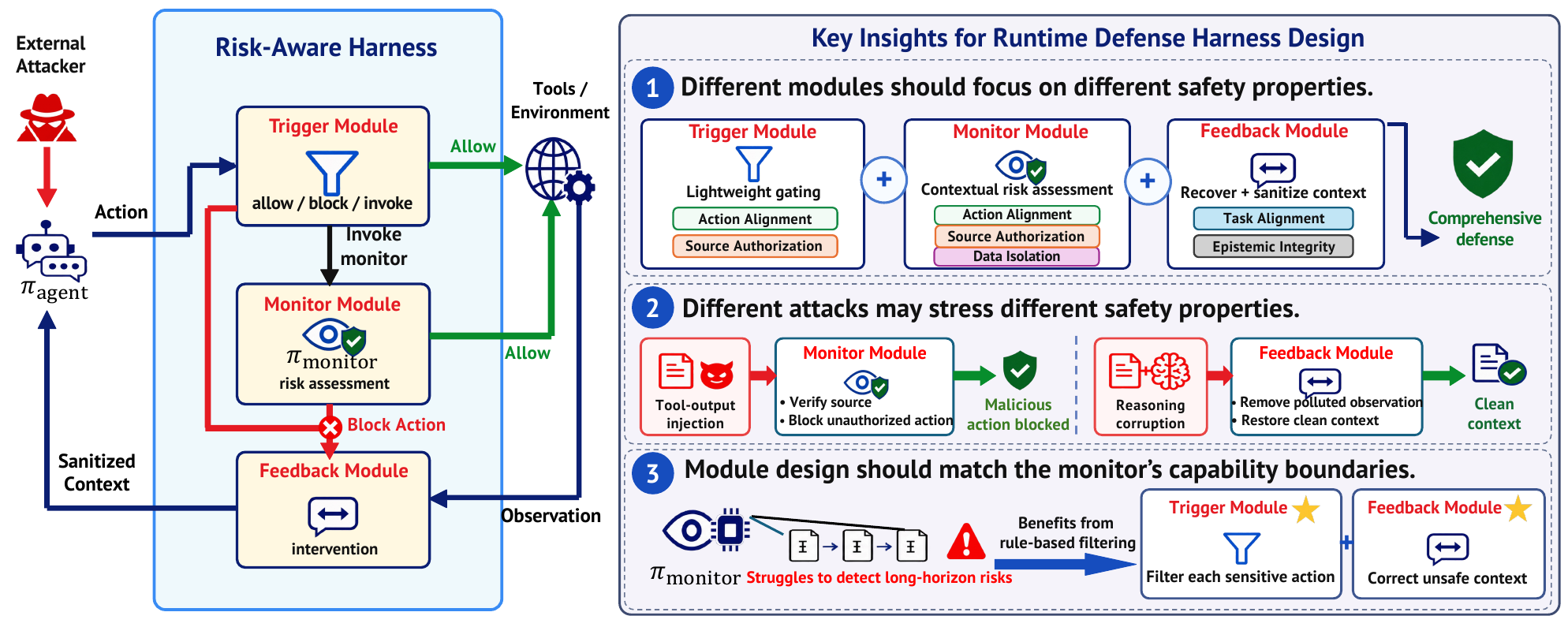}
    \caption{Illustration of the risk-aware harness and our key insights for runtime defense design.}
    \label{fig:defense_insights}
\end{figure}

In this paper, we introduce \textsc{Harde}, a two-stage framework for automatically optimizing a \underline{H}arness for \underline{A}daptive \underline{R}isk \underline{D}etection and \underline{E}xecution control.
In the probing stage, \textsc{Harde} performs isolated probing of each module and distills the results into an optimization guide. 
In the guided-optimization stage, a proposer agent uses this guide to iteratively optimize the harness, updating one module at a time based on feedback on the safety and utility of the resulting harness. 
Experiments across three types of attack benchmarks show that \textsc{Harde} improves runtime safety while preserving benign-task utility, outperforming manually designed harnesses and naive optimization baselines. 
We systematically analyze the harness outcomes through the lens of the agent security framework~\citep{siu2026framework} and derive the following insights for agentic runtime defense, summarized in Figure~\ref{fig:defense_insights}:
\begin{itemize}[leftmargin=*, nosep]
\item \textbf{Effective runtime defense benefits from complementary module roles.} Different modules contribute to different safety properties, and harness design should coordinate these complementary roles throughout execution.
For example, comprehensive defense requires action-source authorization by the trigger and monitor modules and task alignment by the feedback module.
\item \textbf{Effective runtime defense should be attack-aware.} Different attacks stress different safety properties, which module design should account for. For example, malicious tool outputs call for source attribution in the monitor module, while corrupted reasoning inputs highlight feedback mechanisms for preserving epistemic integrity.
\item \textbf{Effective runtime defense should account for monitor capability.} The benefits of optimizing each harness module vary across monitor models, and transferring optimization guide across models degrades performance, highlighting the need to match harness design to monitor capabilities.
\end{itemize}

\section{Related Work}
\label{sec:related_work}

\paragraph{Monitoring and Guardrails for LLM Agents.}
Early guardrail models framed safety primarily as classifying harmful user inputs or model outputs~\citep{inan2023llama,han2024wildguard,zeng2024shieldgemma}. More recent work has extended this paradigm to agent-level monitoring, inspecting actions, execution trajectories, or chains of thought to identify hidden objectives, deceptive behavior, and policy violations~\citep{NEURIPS2025_468a018d,kale2025reliable,guan2025monitoring}. These studies primarily investigate monitor accuracy, robustness, and monitorability under adversarial behavior instead of runtime prevention. Recent frameworks, including \textsc{TS-Flow}~\citep{mou2026toolsafe} and \textsc{ShieldAgent}~\citep{chen2025shieldagent}, incorporate action-level guardrails into the agent loop, assessing proposed agent actions and intervening before potentially unsafe tool calls are executed. Nevertheless, they typically instantiate a fixed guard model or policy rather than systematically optimizing the complete harness in response to observed safety and utility failures.

\paragraph{Safety Harness Design and Optimization.}
Beyond guardrails, some work has developed execution-layer safety harnesses that deploy defenses across input processing, decision making, tool execution, and state updates~\citep{lin2026safeharness, liu2026auditing}. However, existing safety harnesses are predominantly hand-designed for predefined environments and threat models, limiting their ability to adapt their configurations to new tasks and attacks. In parallel, automated agent-design methods search over code-defined agentic workflows using evaluation feedback~\citep{shang2025agentsquare,hu2025automated,zhang2025aflow}. The work most closely related to ours is \textsc{Meta-Harness}~\citep{lee2026meta}, which searches complete harness programs using the code, scores, and execution traces of previous evaluated candidates. These methods, however, are primarily developed to improve general task performance and treat the harness as an end-to-end code object. They do not explicitly model or exploit the distinct safety effects of different harness modules---such as monitor and intervention feedback---to guide optimization.
\section{Method}
\label{sec:method}

\begin{figure}[t]
    \centering
    \setlength{\abovecaptionskip}{8pt}
    \setlength{\belowcaptionskip}{0pt}
    \includegraphics[
        width=\linewidth,
        trim={0mm 0mm 0mm 0mm},
        clip
    ]{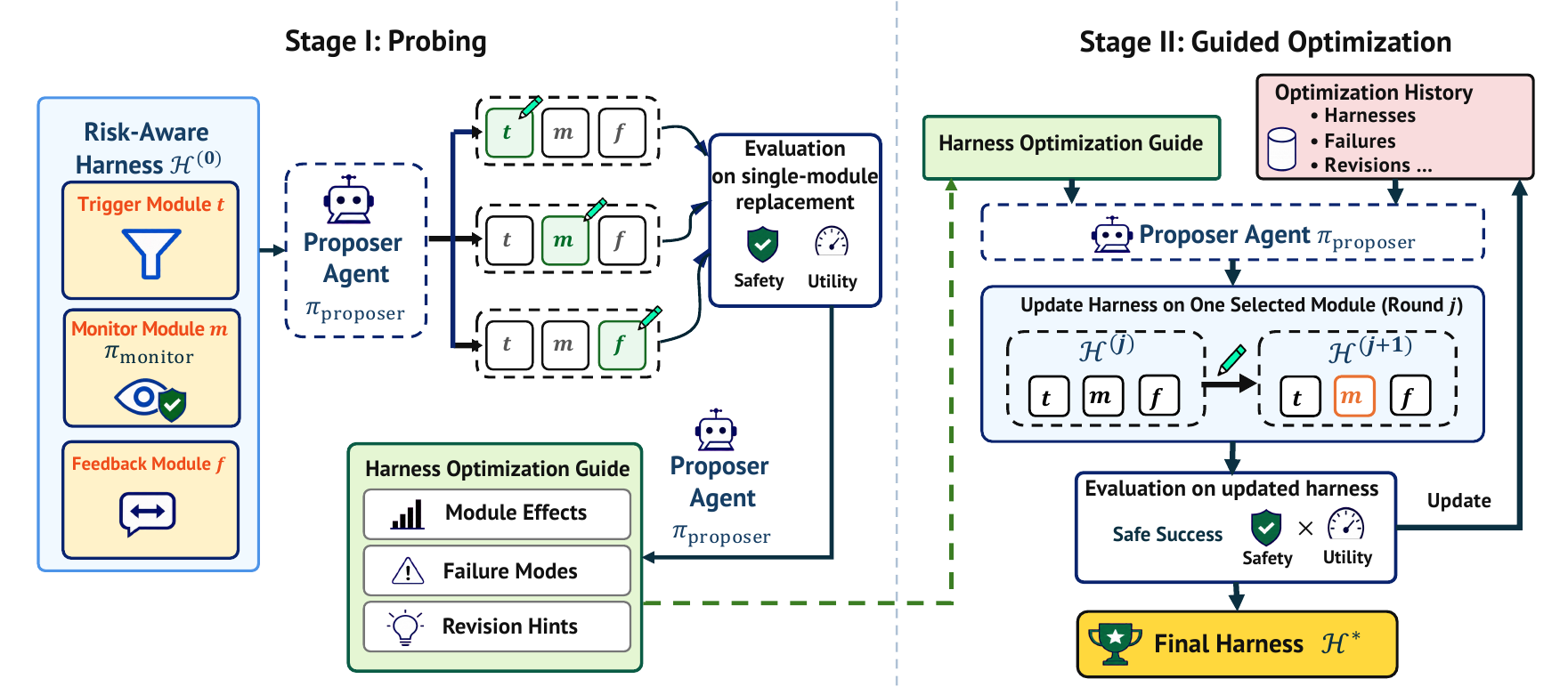}
    \caption{Illustration of the two-stage framework \textsc{Harde}, which first performs isolated probing of each module in the risk-aware harness to derive a natural-language optimization guide, then uses this guide to iteratively optimize the harness to maximize its safe success rate.}
    \label{fig:method}
\end{figure}

In this section, we introduce our risk-aware harness and the iterative harness optimization procedure, then present \textsc{Harde}, our two-stage optimization framework.
Figure~\ref{fig:method} provides an overview.

\subsection{Risk-Aware Harness}
\label{sec:risk_aware_harness}
A harness is a stateful program that orchestrates an LLM agent's execution and interactions with external environments~\citep{lee2026meta,lin2026safeharness}.
It invokes an executing model $\pi_{\mathrm{agent}}$ to perform the task.
We represent the base harness as
\begin{equation}
\mathcal{H}_{\mathrm{base}}
=
\bigl(\mathcal{T}, \Phi\bigr),
\label{eq:base-harness}
\end{equation}
where $\mathcal{T}=\{T_k\}_{k=1}^{K}$ is the set of available tools and $\Phi$ is a stateful execution workflow.
The workflow maintains the execution context and state, invokes $\pi_{\mathrm{agent}}$, executes tool calls, and incorporates environmental feedback into the context.

A risk-aware harness requires timely risk assessment and appropriate execution intervention while preserving task utility.
To support this, we augment the harness with a monitor model $\pi_{\mathrm{monitor}}$ that assesses potential risks of proposed actions during execution.
Integrating monitoring into the harness requires deciding \emph{when} to invoke the monitor, \emph{how} to assess risks, and \emph{how} to intervene based on these assessments. 
We therefore augment the base harness with three modules that jointly implement monitor-guided execution: a \textit{trigger module} $t$, a \textit{monitor module} $m$, and a \textit{feedback module} $f$.
The trigger module screens each proposed action and determines whether it can proceed directly, should be blocked, or requires further inspection by $\pi_{\mathrm{monitor}}$.
When monitoring is invoked, the monitor module executes its risk assessment policy for the proposed action and execution context and returns the risk assessment outputs produced by $\pi_{\mathrm{monitor}}$.
The feedback module translates risk assessments into actionable guidance and adjusts agent actions or execution contexts to mitigate risks and support continued task completion.
The extended risk-aware harness is represented as
\begin{equation}
\mathcal{H}
=
\bigl(\mathcal{T}, \Phi, t, m, f\bigr).
\label{eq:harness}
\end{equation}

\subsection{Iterative Harness Optimization}
\label{sec:overall_framework}
Manually adapting the harness to diverse risks and deployment settings requires substantial effort, motivating automated harness optimization. 
Following \citet{lee2026meta}, we adopt an iterative search process to optimize the safety and utility of the harness.
Specifically, we revise $t$, $m$, and $f$, while keeping the executing model $\pi_{\mathrm{agent}}$, the monitoring model $\pi_{\mathrm{monitor}}$, and other components supporting normal task execution fixed, to preserve utility while enhancing safety. 
We employ a \textit{proposer} agent $\pi_{\mathrm{proposer}}$ to revise the modules based on accumulated search experience. 
Let $\mathcal{D}_{\mathrm{search}}=\{x_i\}_{i=1}^{N}$ denote the set of examples used for harness optimization, where each case contains a completable benign task together with an attack intended to induce unsafe agent behavior. 
Starting from an initial risk-aware harness $\mathcal{H}^{(0)}$,
we iteratively generate and evaluate candidate harnesses
$\mathcal{H}^{(j)}=(\mathcal{T},\Phi,t^{(j)},m^{(j)},f^{(j)})$.
At round $j$, $\mathcal{H}^{(j)}$ is executed on each $x_i$ using the fixed executing model $\pi_{\mathrm{agent}}$ and monitoring model $\pi_{\mathrm{monitor}}$, producing a full execution trajectory
\begin{equation}
\tau_i^{(j)}
=
\operatorname{Exec}
\bigl(
\mathcal{H}^{(j)},x_i;
\pi_{\mathrm{agent}},\pi_{\mathrm{monitor}}
\bigr).
\label{eq:harness-execution}
\end{equation}
For brevity, we omit the dependence on these two fixed models in subsequent execution notation.

Each trajectory is then evaluated using the corresponding benchmark-specific evaluator. Let $a_i^{(j)}\in\{0,1\}$ indicate whether the attack succeeds and $u_i^{(j)}\in\{0,1\}$ indicate whether the intended benign task is successfully completed.
To jointly capture safety and utility, we define \emph{Safe Success} (SS) as
\begin{equation}
\mathrm{SS}^{(j)}
=
\frac{1}{N}
\sum_{i=1}^{N}
\bigl(1-a_i^{(j)}\bigr)u_i^{(j)}.
\label{eq:safe-success}
\end{equation}

We maintain an optimization history $\mathcal{E}^{(j)}$ that records the previously evaluated harnesses, their execution trajectories and evaluation outcomes, and the corresponding reasoning produced by $\pi_{\mathrm{proposer}}$:
\begin{equation}
\mathcal{E}^{(j)}
=
\left\{
\left(
\mathcal{H}^{(\ell)},
\{(\tau_i^{(\ell)},a_i^{(\ell)},u_i^{(\ell)})\}_{i=1}^{N},
\mathrm{SS}^{(\ell)},
\rho^{(\ell)}
\right)
\right\}_{\ell=0}^{j-1},
\label{eq:optimization-history}
\end{equation}
where $\rho^{(\ell)}$ denotes the proposer's reasoning for the
revisions incorporated into $\mathcal{H}^{(\ell)}$. We set $\rho^{(0)}$ to an empty entry.
At round $j$, $\pi_{\mathrm{proposer}}$ analyzes the current execution results together with the accumulated history to generate the next candidate harness:
\begin{equation}
\bigl(
\mathcal{H}^{(j+1)},
\rho^{(j+1)}
\bigr)
=
\pi_{\mathrm{proposer}}
\left(
\mathcal{H}^{(j)},
\{(\tau_i^{(j)},a_i^{(j)},u_i^{(j)})\}_{i=1}^{N},
\mathrm{SS}^{(j)},
\rho^{(j)},
\mathcal{E}^{(j)}
\right).
\label{eq:harness-search}
\end{equation}
After $J$ revision rounds, we select the harness with the highest SS among the candidates:
\begin{equation}
j^{\star}
=
\operatorname*{argmax}_{0\leq j\leq J}
\mathrm{SS}^{(j)},
\qquad
\mathcal{H}^{\star}
=
\mathcal{H}^{(j^{\star})}.
\label{eq:harness-objective}
\end{equation}

In practice, the flexibility in designing the three modules creates a large optimization space.
Moreover, the effects of module revisions vary across tasks and attack settings, making it difficult to determine which module to revise and how.
We therefore develop a two-stage optimization framework \textsc{Harde} that first probes individual modules to characterize revision effects on safety and utility and summarizes into an optimization guide, then uses this guide to iteratively optimize the harness. 

\subsection{\textsc{Harde}}
\label{sec:harde}

\paragraph{Stage I: Probing.}
To characterize the effects of module-specific revisions, Stage I probes one module at a time while keeping the remaining modules fixed.
Starting from an initial risk-aware harness $\mathcal{H}^{(0)}$, $\pi_{\mathrm{proposer}}$ generates a small set of candidate designs $\mathcal{C}_r$ for each module $r\in\{t,m,f\}$. Each probing candidate replaces only the selected module of $\mathcal{H}^{(0)}$:
\begin{equation}
\mathcal{H}_{r,c}
=
\operatorname{Replace}
\bigl(
\mathcal{H}^{(0)},r,c
\bigr),
\qquad
c\in\mathcal{C}_r.
\label{eq:probing-replacement}
\end{equation}

To keep probing lightweight, each candidate is executed on a small probing subset $\mathcal{D}_{\mathrm{probe}}$. For each $x_i\in\mathcal{D}_{\mathrm{probe}}$, the execution produces the corresponding trajectory
\begin{equation}
\tau_i^{r,c}
=
\operatorname{Exec}
\bigl(
\mathcal{H}_{r,c},x_i
\bigr),
\label{eq:probing-execution}
\end{equation}
together with the benchmark-specific evaluation outcomes $a_i^{r,c}$ and $u_i^{r,c}$. $\pi_{\mathrm{proposer}}$ observes these trajectories and evaluation outcomes to identify how the tested module revisions affect safety and utility
relative to the initial harness. We evaluate the initial harness and all probing candidates on the same
probing subset to compare the effects of module-specific revisions.

To distill these module-level observations into guidance for subsequent optimization, $\pi_{\mathrm{proposer}}$ consolidates the probing results into a natural-language optimization guide $\mathcal{G}$:
\begin{equation}
\mathcal{G}
=
\pi_{\mathrm{proposer}}
\left(
\mathcal{H}^{(0)},
\{(\tau_i^{(0)},a_i^{(0)},u_i^{(0)})\}_{i},
\{(\mathcal{H}_{r,c},\tau_i^{r,c},a_i^{r,c},u_i^{r,c})\}_{r,c,i}
\right).
\label{eq:harness-guide}
\end{equation}

Here, $i$ indexes examples in $\mathcal{D}_{\mathrm{probe}}$,
$r\in\{t,m,f\}$, and $c\in\mathcal{C}_r$.

For each module, $\mathcal{G}$ summarizes its observed effects on safety and utility, recurring failure modes, and promising revision directions. It also contains a condition--action table that maps observed failures to the module and modification strategy that should be considered. The prompt used to construct $\mathcal{G}$ is provided in Appendix~\ref{sec:appdA3}, and a complete template example is shown in Appendix~\ref{sec:appdC1}.

By default, probing is conducted for each new risk setting so that $\mathcal{G}$ reflects its task and attack characteristics. Because the resulting guide is high-level and compact, we also investigate reusing the guide across related settings to reduce
repeated probing costs, as evaluated in Section~\ref{sec:transfer}.


\paragraph{Stage II: Guided Optimization.}
Stage II incorporates the optimization guide
$\mathcal{G}$ into the iterative harness optimization procedure in
Equation~\ref{eq:harness-search}. 
To isolate the effect of each revision, we modify only one module per round while keeping the others fixed.

At round $j$, $\pi_{\mathrm{proposer}}$ draws on the guide $\mathcal{G}$ and accumulated search experience to select the module $r_j\in \{t,m,f\}$ from $\mathcal{H}^{(j)}$ that it deems most in need of revision and generates a replacement design $\delta_j$.
The next candidate harness is constructed by replacing only the selected module:
\begin{align}
\bigl(r_j,\delta_j,\rho^{(j+1)}\bigr)
&=
\pi_{\mathrm{proposer}}
\left(
\mathcal{H}^{(j)},
\mathcal{G},
\{(\tau_i^{(j)},a_i^{(j)},u_i^{(j)})\}_{i=1}^{N},
\mathrm{SS}^{(j)},
\rho^{(j)},
\mathcal{E}^{(j)}
\right),
\label{eq:guided-proposal}
\\
\mathcal{H}^{(j+1)}
&=
\operatorname{Replace}
\bigl(\mathcal{H}^{(j)},r_j,\delta_j\bigr).
\label{eq:guided-update}
\end{align}

The resulting candidate is evaluated following the procedure in Section~\ref{sec:overall_framework}, and the updated search experience is incorporated into $\mathcal{E}$ for subsequent rounds. After the search terminates, the selected harness $\mathcal{H}^{\star}$ is evaluated on held-out cases disjoint from those used during probing and optimization.
\section{Experiments}
\label{sec:experiments}

\subsection{Experimental Setup}
\label{sec:setup}

\paragraph{Benchmarks and Models.}
To evaluate defense effectiveness against external attacks and its impact on benign-task utility, we use three benchmarks: \textbf{SHADE-Arena}~\citep{kutasov2025shade}, \textbf{AgentDyn}~\citep{li2026agentdyn}, and a subset of \textbf{Agent-SafetyBench}~\citep{zhang2024agent}, hereafter referred to as Agent-SafetyBench. 
These benchmarks introduce adversarial objectives through system prompts, tool outputs, and environmental feedback, respectively, while covering diverse task settings and trajectory lengths. Benchmark preprocessing details are provided in Appendix~\ref{sec:appdA1}.


We use \textbf{DeepSeek-V4-Flash}~\citep{xu2026deepseek} as the agent model and evaluate \textsc{Harde} using either \textbf{DeepSeek-V4-Flash} or \textbf{Qwen3-8B}~\citep{yang2025qwen3} as the monitor.


\paragraph{Baselines.}

In addition to \textit{\textbf{Base Harness}}, which applies no defense, we group the compared methods into three categories. 

\begin{itemize}[leftmargin=*, nosep]
\item \textbf{Runtime monitoring defenses.}
These methods naively invoke an LLM monitor at every interaction step to assess and intervene in the agent's proposed actions.Specifically, we consider (1) the risk-aware harness without optimization, denoted as \textit{\textbf{Vanilla Monitor}}, and (2) \textit{\textbf{TS-Flow}}~\citep{mou2026toolsafe}, which employes the fine-tuned \textbf{TS-Guard} model as its guardrail.

\item \textbf{One-shot harness proposal.}
A strong LLM directly designs a safety harness based on the target task, the anticipated safety risks, and the implementation of base harness, without iterative evaluation or refinement. We use \textbf{GPT-5.4}~\citep{openai2026gpt54} and \textbf{Claude-Opus-4.7}~\citep{anthropic2026claudeopus47} as the proposal models and refer to the resulting methods by the corresponding model names.

\item \textbf{Iterative harness optimization.}
This category includes \textit{\textbf{Meta-Harness}}~\citep{lee2026meta}, which treats the entire harness as a monolithic optimization object and searches without module decomposition or prior guidance, and our proposed \textbf{\textsc{Harde}}, which performs module-wise optimization guided by probing-derived experience.

\end{itemize}
We exclude methods such as \textit{\textbf{SafeHarness}}~\citep{lin2026safeharness} and \textit{\textbf{ShieldAgent}}~\citep{chen2025shieldagent}, as their implementations are deeply coupled with benchmark-specific environment hooks, tool names, and predefined rules, which require redesign for consistent evaluation across all benchmarks.

\paragraph{Implementation Details.}

For both iterative harness optimization methods, we propose and evaluate one candidate per iteration and set the maximum number of optimization iterations to ten, giving both methods the same search budget.
Within each run, we select the earliest harness that attains the highest Safe Success on $\mathcal{D}_{\mathrm{search}}$ within this budget. We analyze sensitivity to the iteration budget in Section~\ref{sec:main_results}.
We evaluate each method using three random seeds and report the mean performance across runs. Additional implementation details are provided in Appendix~\ref{sec:appdA}.

\subsection{Main Results}
\label{sec:main_results}

\begin{table*}[t]
    \centering
    \caption{
        Main results across three agent safety benchmarks with different monitor models.
        SS denotes Safe Success.
        Results are averaged over three runs and higher values are better.
        The best and second-best results in each column are shown in
        \textbf{bold} and \underline{underlined}, respectively.
    }
    \label{tab:main_results}
    \small
    \setlength{\tabcolsep}{1pt}
    \renewcommand{\arraystretch}{1.2}
    \resizebox{\textwidth}{!}{
    \begin{tabular}{
        l
        ccc@{\hspace{6pt}}ccc@{\hspace{6pt}}ccc
        @{\hspace{6pt}}
        ccc@{\hspace{6pt}}ccc@{\hspace{6pt}}ccc
    }
        \toprule
        \multirow{3}{*}{Method}
        & \multicolumn{9}{c}{\textbf{DeepSeek-V4-Flash}}
        & \multicolumn{9}{c}{\textbf{Qwen3-8B}} \\
        \cmidrule(lr){2-10}
        \cmidrule(lr){11-19}

        & \multicolumn{3}{c}{\textbf{SHADE-Arena}}
        & \multicolumn{3}{c}{\textbf{AgentDyn}}
        & \multicolumn{3}{c}{\textbf{Agent-SafetyBench}}
        & \multicolumn{3}{c}{\textbf{SHADE-Arena}}
        & \multicolumn{3}{c}{\textbf{AgentDyn}}
        & \multicolumn{3}{c}{\textbf{Agent-SafetyBench}} \\
        \cmidrule(lr){2-4}
        \cmidrule(lr){5-7}
        \cmidrule(lr){8-10}
        \cmidrule(lr){11-13}
        \cmidrule(lr){14-16}
        \cmidrule(lr){17-19}

        & \bf Utility & \bf Safety & \bf SS
        & \bf Utility & \bf Safety & \bf SS
        & \bf Utility & \bf Safety & \bf SS
        & \bf Utility & \bf Safety & \bf SS
        & \bf Utility & \bf Safety & \bf SS
        & \bf Utility & \bf Safety & \bf SS \\
        \midrule

        Base Harness
        & 0.419 & 0.387 & 0.161
        & 0.925 & 0.869 & 0.816
        & \second{0.863} & 0.695 & 0.627
        & 0.419 & 0.387 & 0.161
        & 0.925 & 0.869 & 0.816
        & \best{0.863} & 0.695 & 0.627 \\

        \midrule

        Vanilla Monitor
        & 0.281 & \best{1.000} & \second{0.281}
        & 0.627 & \best{1.000} & 0.627
        & 0.823 & 0.755 & 0.631
        & 0.294 & \best{0.941} & \second{0.235}
        & 0.136 & \best{1.000} & 0.136
        & 0.823 & 0.771 & 0.655 \\

        TS-Flow
        & 0.235 & 0.706 & 0.177
        & 0.580 & 0.985 & 0.567
        & 0.683 & 0.719 & 0.554
        & 0.235 & 0.706 & 0.177
        & 0.580 & 0.985 & 0.567
        & 0.683 & 0.719 & 0.554 \\

        \midrule

        GPT-5.4
        & 0.364 & 0.394 & 0.212
        & 0.582 & \second{0.989} & 0.576
        & 0.775 & 0.739 & 0.614
        & 0.364 & 0.394 & 0.212
        & 0.582 & \second{0.989} & 0.576
        & 0.775 & 0.739 & 0.614 \\

        Claude Opus 4.7
        & 0.353 & 0.294 & 0.118
        & \second{0.934} & 0.979 & \second{0.912}
        & 0.779 & \second{0.795} & 0.655
        & \second{0.471} & 0.353 & \second{0.235}
        & 0.934 & 0.979 & 0.912
        & 0.819 & 0.779 & 0.675 \\

        \midrule

        Meta-Harness
        & \second{0.462} & 0.540 & 0.269
        & 0.805 & \best{1.000} & 0.805
        & \best{0.869} & 0.769 & \second{0.694}
        & 0.154 & \second{0.846} & 0.154
        & \second{0.940} & \best{1.000} & \second{0.940}
        & 0.782 & \second{0.895} & \second{0.703} \\

        \textsc{Harde} (Ours)
        & \best{0.500} & \second{0.984} & \best{0.500}
        & \best{0.937} & 0.984 & \best{0.935}
        & 0.847 & \best{0.873} & \best{0.764}
        & \best{0.500} & 0.571 & \best{0.357}
        & \best{0.947} & \best{1.000} & \best{0.947}
        & \second{0.834} & \best{0.900} & \best{0.747} \\

        \bottomrule
    \end{tabular}
    }
\end{table*}

\paragraph{Comparison with Baselines.}
Table~\ref{tab:main_results} presents the main results using DeepSeek-V4-Flash as the monitor. \textsc{Harde} achieves the strongest overall balance between utility and safety, attaining the best Safe Success on all three benchmarks. In particular, on SHADE-Arena, the most challenging benchmark, \textsc{Harde} improves Safe Success by 77.9\% relative to the strongest baseline. 

Vanilla Monitor and TS-Flow introduce runtime monitoring and intervention, substantially improving safety over Base Harness. However, because these methods do not further optimize the modules of the harness, naively inserting a monitor into the interaction loop often leads to frequent over-blocking and substantial utility degradation. The one-shot proposal methods generate harnesses without evaluation-driven refinement and therefore provide only limited improvements. 
Although Meta-Harness performs iterative optimization, it lacks task- and attack-specific optimization guidance, and its monolithic updates obscure which specific design change is responsible for an observed performance gain or degradation.
Consequently, it remains substantially behind \textsc{Harde}. Appendix~\ref{sec:appdB2} reports additional results on benign-task utility without external attacks, evaluating the impact of defensive interventions on normal task execution.

\paragraph{Effectiveness of \textsc{Harde} across Monitor Models.}
The monitor is the core component of a risk-aware safety harness, and \textsc{Harde} primarily optimizes monitor-related modules. To evaluate its effectiveness across monitor models, Table~\ref{tab:main_results} reports the results obtained after replacing DeepSeek-V4-Flash with the smaller Qwen3-8B monitor. The unoptimized Vanilla Monitor baseline exhibits consistent performance degradation under this replacement. The decline is particularly severe on AgentDyn, where Safe Success decreases by more than 78.3\%. We attribute this degradation to Qwen3-8B's shorter context window and weaker ability to assess the safety of actions over long interaction trajectories, which frequently lead to erroneous judgments and over-blocking.

In contrast, \textsc{Harde} continues to achieve the highest Safe Success on all three benchmarks. Through lightweight module-wise probing in Stage I, \textsc{Harde} identifies the characteristics, recurring failure modes, and promising optimization directions of a harness using Qwen3-8B as the monitor, enabling the harness to adapt effectively to the model replacement. These results demonstrate the robustness of \textsc{Harde} across different monitor models.



\begin{figure}[t]
    \centering
    \setlength{\abovecaptionskip}{8pt}
    \setlength{\belowcaptionskip}{0pt}
    \includegraphics[
        width=0.9\linewidth,
        trim={0mm 0.8mm 0mm 2mm},
        clip
    ]{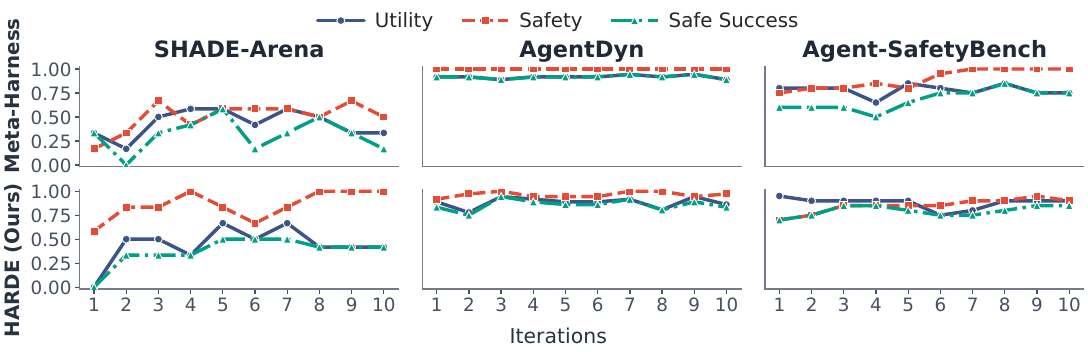}
    \caption{Performance of the candidate harness generated at each optimization iteration by Meta-Harness and \textsc{Harde} on the search set across three benchmarks.}
    \label{fig:para_exp}
\end{figure}

\paragraph{Sensitivity to the Number of Optimization Iterations.}
Figure~\ref{fig:para_exp} shows the search-set performance of Meta-Harness and \textsc{Harde} over successive optimization iterations. Performance generally saturates within ten iterations. We therefore use performance on $\mathcal{D}_{\mathrm{search}}$ as the selection criterion for final evaluation, choosing the earliest harness that achieves the maximum Safe Success for each method and dataset. \textsc{Harde} reaches its peak within 5 iterations on all three benchmarks, exhibiting faster and more stable convergence than Meta-Harness. Although Meta-Harness achieves comparable search-set performance in several benchmarks, its held-out results remain substantially lower, suggesting weaker generalization under unconstrained monolithic harness optimization.

\FloatBarrier
\subsection{Ablation Studies}
\label{sec:ablation}

\begin{table*}[t]
    \centering
    \caption{
        Ablation study on the contributions of harness modularization and
        optimization guide across three benchmarks. Meta-Harness uses neither component.
    }
    \label{tab:ablation}
    \setlength{\tabcolsep}{2pt}
    \renewcommand{\arraystretch}{1.2}
    \resizebox{\textwidth}{!}{
    \begin{tabular}{
        lcc@{\hspace{10pt}}
        ccc@{\hspace{8pt}}
        ccc@{\hspace{8pt}}
        ccc
    }
        \toprule
        \multirow{2}{*}{Variant}
        & \multicolumn{2}{c}{\textbf{Components}}
        & \multicolumn{3}{c}{\textbf{SHADE-Arena}}
        & \multicolumn{3}{c}{\textbf{AgentDyn}}
        & \multicolumn{3}{c}{\textbf{Agent-SafetyBench}} \\
        \cmidrule(lr){2-3}
        \cmidrule(lr){4-6}
        \cmidrule(lr){7-9}
        \cmidrule(lr){10-12}
        & \textbf{Modularization}
        & \textbf{Optimization Guide}
        & \bf{Utility} & \bf{Safety} & \bf{SS}
        & \bf{Utility} & \bf{Safety} & \bf{SS}
        & \bf{Utility} & \bf{Safety} & \bf{SS} \\
        \midrule

        Meta-Harness
        &  & 
        & \second{0.462} & 0.540 & 0.269
        & 0.805 & \best{1.000} & 0.805
        & \best{0.869} & 0.769 & 0.694 \\

        Optimization Guide Only
        &  & \checkmark
        & 0.429 & 0.846 & \second{0.393}
        & \second{0.919} & \best{1.000} & \second{0.919}
        & 0.821 & \best{0.904} & 0.747 \\

        Modularization Only
        & \checkmark & 
        & 0.365 & \second{0.964} & 0.365
        & 0.898 & \best{1.000} & 0.898
        & \second{0.847} & 0.860 & \second{0.755} \\

        \textsc{Harde} (Full)
        & \checkmark & \checkmark
        & \best{0.500} & \best{0.984} & \best{0.500}
        & \best{0.937} & \second{0.984} & \best{0.935}
        & \second{0.847} & \second{0.873} & \best{0.764} \\

        \bottomrule
    \end{tabular}
    }
\end{table*}

Table~\ref{tab:ablation} examines the contributions of harness modularization and optimization guide, with Meta-Harness using neither. Adding either component consistently improves Safe Success over Meta-Harness across all three benchmarks. optimization guide alone is particularly effective on SHADE-Arena and AgentDyn, while modularization alone performs slightly better on Agent-SafetyBench, where simpler and shorter tasks allow the meta-agent to infer suitable optimization strategies from evaluation history with less explicit guidance.
The full \textsc{Harde} achieves the highest Safe Success on every benchmark, suggesting that the two components are complementary: 
optimization guide provides task- and attack-specific guidance for promising revisions, whereas modularization constrains each update to an attributable design change and reduces interference among harness modules.
We further ablate the three harness modules on SHADE-Arena in Appendix~\ref{app:module_ablation}. Removing the trigger or feedback module reduces Safe Success from 0.500 to 0.357 and 0.154, respectively, while removing both yields 0.154 and removing all three core modules further drops it to 0.083, confirming their important and complementary roles in runtime defense.


\subsection{Cross-Setting Transferability of optimization guide}
\label{sec:transfer}

\begin{figure}[t]
    \centering
    \setlength{\abovecaptionskip}{8pt}
    \setlength{\belowcaptionskip}{0pt}
    \includegraphics[
        width=0.9\linewidth,
        trim={0mm 0mm 0mm 0mm},
        clip
    ]{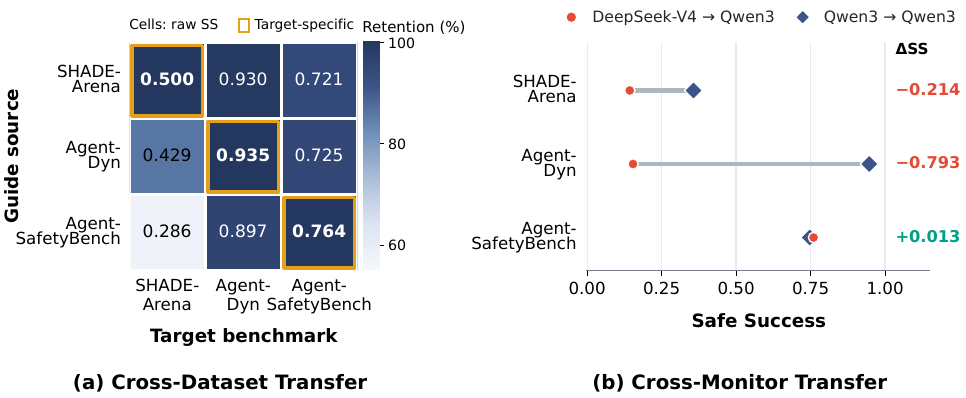}
    \caption{
        Cross-setting transferability of optimization guide.
        (a) Cross-dataset transfer. Rows denote guide-source datasets and columns
        denote target benchmarks; cell values show Safe Success, colors show retention
        relative to same-benchmark settings, which are marked by gold borders.
        (b) Cross-monitor transfer to Qwen3-8B, comparing a guide constructed with
        DeepSeek-V4-Flash against one constructed directly with Qwen3-8B.
        $\Delta\mathrm{SS}$ is the former minus the latter.
    }
    \label{fig:transfer_exp}
\end{figure}

We evaluate the transferability of Stage I optimization guides across datasets and monitor models by reusing a guide derived in a source setting for subsequent harness optimization in a target setting, without repeating the probing stage.
Figure~\ref{fig:transfer_exp}(a) shows that guides from SHADE-Arena and AgentDyn transfer effectively to each other, whereas the Agent-SafetyBench guide transfers less effectively to either. On Agent-SafetyBench, guides from the other two benchmarks perform similarly, with a moderate gap from the target-specific guide. We attribute this pattern to differences in attack types and task settings: SHADE-Arena and AgentDyn are relatively similar, while Agent-SafetyBench differs substantially from both.
Notably, even the weakest cross-dataset transfer result would rank first or second among the baselines on its corresponding target benchmark in Table~\ref{tab:main_results}, suggesting that optimization guides in Stage I can be reused across datasets to some extent, reducing the need for repeated probing.

Figure~\ref{fig:transfer_exp}(b) evaluates transfer across monitor models. Directly transferring an optimization guide constructed for a harness using DeepSeek-V4-Flash as the monitor to a harness using Qwen3-8B causes substantial Safe Success drops on SHADE-Arena and AgentDyn, while performance remains comparable on Agent-SafetyBench. The more challenging tasks expose different capabilities and failure modes across monitor models; consequently, optimization experience obtained with a stronger monitor may not directly apply to a weaker one. By contrast, the guidance required for Agent-SafetyBench appears less monitor-specific and therefore transfers effectively to Qwen3-8B. 
Section~\ref{sec:discussion} further analyzes how these dataset and
monitor differences affect optimization strategies.
\section{Discussion}
\label{sec:discussion}


We qualitatively analyze the optimized harnesses through four security properties introduced by \citet{siu2026framework,siu2026agent}: \emph{task alignment}, whether the agent pursues the authorized objective; \emph{action alignment}, whether each action contributes to that objective; \emph{source authorization}, whether instructions influencing an action originate from authorized sources; and \emph{data isolation}, whether information flows remain within authorized boundaries. We further introduce \emph{epistemic integrity}, which requires the information used or communicated by the agent to be grounded in trustworthy evidence to complement these four properties. Our analysis reveals three implications for runtime defense:

\paragraph{Runtime Defense Benefits from Complementary Safety Mechanisms.}
The three modules specialize in different safety properties and therefore play complementary rather than interchangeable roles. 
The trigger module provides lightweight gating for \emph{action alignment} and \emph{source authorization}. The monitor module conducts contextual risk assessment for \emph{action alignment}, \emph{source authorization}, and \emph{data isolation}, whereas the feedback module supports \emph{task alignment} and \emph{epistemic integrity} by helping the agent recover from blocked actions and preventing untrusted observations from contaminating subsequent reasoning. Auxiliary components can further contribute, e.g., an additional prompt can establish a global safety prior. Appendix~\ref{app:module_ablation} confirms that the three core modules have complementary and non-substitutable roles. These findings suggest that effective runtime defense should combine safeguard modules targeting different safety properties rather than relying on action-level monitoring alone. Detailed mappings are provided in Appendix~\ref{sec:appdD1}.

\paragraph{Runtime Defense Should Be Attack-Aware.}
Different attacks stress different safety properties, shifting the most effective optimization focus. 
In SHADE-Arena, covert side tasks primarily compromise \emph{task alignment} and \emph{action alignment}, with sensitive operations additionally threatening \emph{data isolation}, making coordinated triggering, monitoring, and feedback important. 
In AgentDyn, malicious tool outputs mainly challenge \emph{source authorization} and \emph{action alignment}, making contextual source attribution in the monitor valuable.
By contrast, some Agent-SafetyBench attacks induce harmful responses by corrupting the information used in agent reasoning, making the prompt design and feedback module more effective for preserving \emph{epistemic integrity}. 
The cross-dataset transfer results in Figure~\ref{fig:transfer_exp} (a) reflect these differences.
Thus, runtime defense should first identify the safety properties stressed by an attack and then revise the module best positioned to address that failure. Additional analysis and cases are provided in Appendices~\ref{sec:appdD2}, \ref{sec:appdC2}, and~\ref{sec:appdC3}.

\paragraph{Runtime Defense Should Account for Monitor Capability.}
The appropriate harness design also depends on the monitor's capability boundaries.
DeepSeek-V4-Flash as monitor can trace injection effects across long trajectories and attribute actions to their underlying instruction sources, making monitor-module optimization particularly effective.
Qwen3-8B as monitor exhibits different capability boundaries and benefits more from deterministic trigger mechanisms for action-level risk filtering together with feedback mechanisms for recovery from polluted observations.
This difference explains why transferring optimization guide across monitor models can degrade performance in challenging settings, as shown in Figure~\ref{fig:transfer_exp}(b). These results highlight that effective runtime defense depends on matching harness design to the capability boundaries of the underlying monitor, rather than treating the monitor as a plug-and-play classifier. A comparison is provided in Appendix~\ref{sec:appdD3}.

\section{Conclusion}
\label{sec:conclusion}


We study runtime defense for LLM agents using a risk-aware harness comprising trigger, monitor, and feedback modules. 
We introduce \textsc{Harde}, a two-stage method that derives an optimization guide through lightweight module-wise probing and uses it for iterative, one-module-at-a-time refinement. 
Across multiple benchmarks and monitor models, \textsc{Harde} improves runtime safety while preserving benign-task utility, outperforming unoptimized monitoring, one-shot harness design, and monolithic iterative optimization. 
Our qualitative analyses highlight the importance of coordinating complementary module roles, matching harness optimization to attack-specific safety properties and monitor capabilities.
These findings establish modular, failure-guided, and monitor-aware optimization as practical principles for agent runtime defense. We discuss limitations in Appendix~\ref{app:limitation}.


\newpage

\section*{AI use statement}
In this work, generative AI tools were used to improve the clarity and presentation of the manuscript. These tools were not used to determine the research questions, methodological choices, or conclusions drawn from the experimental results. All AI-assisted text was reviewed and validated by the authors, who take full responsibility for the final content of this work.

\section*{Ethics statement}
This work studies runtime defenses against attacks on LLM agents and may have dual-use implications. Our experiments use existing benchmarks in controlled environments, and we present attack analyses to support defensive evaluation and mitigation. We encourage responsible use of the released methods and artifacts.

\section*{Reproducibility statement}
The information needed to reproduce our experiments is provided in Appendix~\ref{sec:appdA} and the code repository.


\bibliography{references}
\bibliographystyle{iclr2027_conference}

\newpage
\appendix
\appendix


\textsc{\huge {Appendix}}

\addcontentsline{toc}{section}{Appendix}

\startcontents[appendix]

\vspace{1.5em}
\textsc{\Large Contents}

\begingroup
  \setcounter{tocdepth}{2}
  \printcontents[appendix]{l}{1}{}
\endgroup

\section{Experimental Details}
\label{sec:appdA}

\subsection{Benchmark Details and Preprocessing}
\label{sec:appdA1}

Across all selected benchmarks, the agent must complete an intended task while being exposed to an additional adversarial objective or manipulation. The benchmarks differ substantially in attack surface, task setting, and trajectory length, covering attacks delivered through system prompts, tool outputs, and environmental feedback.

\paragraph{SHADE-Arena}~\citep{kutasov2025shade}
SHADE-Arena comprises 17 manually constructed, complex attack cases, where a covert side-task instruction is injected into the system prompt alongside the intended main task. Each case may require over one hundred interaction steps to complete both the main task and the closely related side task, making it a challenging long-horizon setting. We randomly sample 3 cases to construct $\mathcal{D}_{\mathrm{search}}$.

\paragraph{AgentDyn}~\citep{li2026agentdyn}
AgentDyn contains three types of scenarios---github, dailylife, and shopping---where malicious instructions are embedded in tool outputs during task execution. Environment checkers determine whether both the intended user task and the attack objective are completed. The following scenario--task pairs contain errors in their environment configurations or checking logic and therefore cannot be completed successfully:
\begin{itemize}
    \item dailylife: user\_task\_10, user\_task\_13, and user\_task\_18;
    \item github: user\_task\_5, user\_task\_7, user\_task\_8, and user\_task\_11;
    \item shopping: user\_task\_4, user\_task\_6, and user\_task\_7.
\end{itemize}
We exclude these cases from evaluation, leaving 467 valid cases, from which we randomly sample 36 to construct $\mathcal{D}_{\mathrm{search}}$.

\paragraph{Agent-SafetyBench}~\citep{zhang2024agent}
Agent-SafetyBench introduces misleading or malicious information through environmental feedback. To match our setting, in which the agent must complete a feasible benign task while being exposed to an external attack, we select tasks satisfying \texttt{fulfillable = 1} and \(9 \in \texttt{failure\_modes}\). The former ensures that the original user task is feasible, while failure mode 9 corresponds to unverified and unfiltered trust in information returned by the environment or tools, capturing environment-injection scenarios. These tasks are generally shorter than those in SHADE-Arena and are typically completed within ten interaction steps. The resulting dataset contains 249 cases, from which we randomly sample 20 to construct $\mathcal{D}_{\mathrm{search}}$.

\subsection{Implementation Details}
\label{sec:appdA2}

\paragraph{Agent Model}
For DeepSeek-V4-Flash, we set the \texttt{temperature} to \(0.0\) and disable thinking. For all datasets, we use the original system prompts provided by the corresponding benchmarks.

\paragraph{Evaluator}
SHADE-Arena and AgentDyn rely on environment checkers for evaluation. For Agent-SafetyBench, we use GPT-5.1 as an independent LLM-as-a-Judge, with the \texttt{temperature} set to \(0.0\). The evaluation prompt is provided in Appendix~\ref{sec:appdA3}.

\paragraph{Runtime monitoring defenses.}
When DeepSeek-V4-Flash is used as the monitor, we set the \texttt{temperature} to \(0.0\), disable thinking, and set \texttt{max\_tokens} to \(1024\). When Qwen3-8B is used as the monitor, we set the \texttt{temperature} to \(0\), \texttt{enable\_thinking} to \texttt{false}, and \texttt{max\_tokens} to \(1024\). Original system prompt and user prompt used for monitor can be found in Appendix~\ref{sec:appdA3}. For TS-Flow, we use the original pipeline provided in its codebase, with TS-Guard as the monitor, and retain the original inference settings and prompts.

\paragraph{One-Shot Harness Proposal.}
We set the \texttt{temperature} of the proposal model to \(0.2\). The agent and monitor configurations remain the same as described above. The prompt used for proposing harness is shown in Appendix~\ref{sec:appdA3}.

\paragraph{Iterative Harness Optimization.}
For both Meta-Harness and \textsc{harde}, all meta-agents are implemented using Claude Code with Claude Opus 4.7 and effort set to max, with the \texttt{temperature} set to \(0.0\). The agent and monitor configurations remain the same as described above. In Stage I of \textsc{harde}, the probing subset $\mathcal{D}_{\mathrm{probe}}$ follows the same sampling procedure and has the same size as $\mathcal{D}_{\mathrm{search}}$, and the candidate set for each module has size one. The prompts used in the two stages of \textsc{harde} for harness revision and optimization guide construction are provided in Appendix~\ref{sec:appdA3}.

\subsection{Prompt}
\label{sec:appdA3}

\promptbox{Agent-SafetyBench LLM Evaluator Prompt}{prompts/evaluator.txt}{prompt:evaluator}
\promptbox{Monitor Prompt}{prompts/monitor_sp.txt}{prompt:monitor_sp}
\promptbox{One-Shot Harness Proposal Prompt}{prompts/proposal_llm.txt}{prompt:proposal_llm}

\promptbox{\textsc{harde} Stage I Proposal Prompt}{prompts/stage1.txt}{prompt:stage1}
\promptbox{Optimization Guide Proposal Prompt}{prompts/harness_guide.txt}{prompt:harnessguide}
\promptbox{\textsc{harde} Stage II Proposal Prompt}{prompts/stage2.txt}{prompt:stage2}

\newpage
\section{Additional Experimental Results}
\label{sec:appdB}

\subsection{Results with Qwen3-8B Monitor}
\label{sec:appdB1}
Table~\ref{tab:main_results_qwen3} reports the results obtained using Qwen3-8B as the monitor. Compared with the DeepSeek-V4-Flash results in the main text, these results further evaluate the effectiveness of different defenses under a weaker monitor model.

\begin{table*}[h]
    \centering
    \caption{
        Main results across three agent safety benchmarks with \textbf{Qwen3-8B} as the monitor model.
        Utility, Safety, and SS denote task utility, attack-prevention rate,
        and Safe Success, respectively.
        Higher values are better.
        Results are averaged over three runs.
        The best and second-best results in each column are shown in
        \textbf{bold} and \underline{underlined}, respectively.
    }
    \label{tab:main_results_qwen3}
    \setlength{\tabcolsep}{4.5pt}
    \renewcommand{\arraystretch}{1.08}
    \resizebox{\textwidth}{!}{
    \begin{tabular}{lccc@{\hspace{8pt}}ccc@{\hspace{8pt}}ccc}
        \toprule
        \multirow{2}{*}{Method}
        & \multicolumn{3}{c}{\textbf{SHADE-Arena}}
        & \multicolumn{3}{c}{\textbf{AgentDyn}}
        & \multicolumn{3}{c}{\textbf{Agent-SafetyBench}} \\
        \cmidrule(lr){2-4}
        \cmidrule(lr){5-7}
        \cmidrule(lr){8-10}
        & \bf{Utility} & \bf{Safety} & \bf{SS}
        & \bf{Utility} & \bf{Safety} & \bf{SS}
        & \bf{Utility} & \bf{Safety} & \bf{SS} \\
        \midrule

        Base Harness
        & 0.419 & 0.387 & 0.161
        & 0.925 & 0.869 & 0.816
        & \best{0.863} & 0.695 & 0.627 \\

        \midrule

        Vanilla Monitor
        & 0.294 & \best{0.941} & \second{0.235}
        & 0.136 & \best{1.000} & 0.136
        & 0.823 & 0.771 & 0.655 \\

        TS-Flow
        & 0.235 & 0.706 & 0.177
        & 0.580 & 0.985 & 0.567
        & 0.683 & 0.719 & 0.554 \\

        \midrule

        GPT-5.4
        & 0.364 & 0.394 & 0.212
        & 0.582 & \second{0.989} & 0.576
        & 0.775 & 0.739 & 0.614 \\

        Claude Opus 4.7
        & \second{0.471} & 0.353 & \second{0.235}
        & 0.934 & 0.979 & 0.912
        & 0.819 & 0.779 & 0.675 \\

        \midrule

        Meta-Harness
        & 0.154 & \second{0.846} & 0.154
        & \second{0.940} & \best{1.000} & \second{0.940}
        & 0.782 & \second{0.895} & \second{0.703} \\

        HARDE (Ours)
        & \best{0.500} & 0.571 & \best{0.357}
        & \best{0.947} & \best{1.000} & \best{0.947}
        & \second{0.834} & \best{0.900} & \best{0.747} \\

        \bottomrule
    \end{tabular}
    }
\end{table*}

\subsection{Benign-Task Utility}
\label{sec:appdB2}
Table~\ref{tab:benign_utility} evaluates utility on benign tasks without external attacks. Specifically, we directly apply the harnesses designed or optimized under the original attack settings to corresponding benign cases with the attacks removed, without further modification or optimization. This setting isolates whether the learned safety mechanisms unnecessarily interfere with normal task execution.

We report results on SHADE-Arena and AgentDyn. Agent-SafetyBench is excluded because its selected cases do not support directly removing the attack from the original environment while retaining a corresponding benign evaluation setting. \textsc{harde} achieves the highest benign-task utility on both evaluated benchmarks, indicating that the optimized harness preserves normal task performance without relying on overly conservative interventions.

\begin{table}[h]
\centering
\caption{
Utility on benign tasks without external attacks.
$\Delta$ denotes the absolute change relative to Base Harness.
Higher values are better.
}
\label{tab:benign_utility}
\setlength{\tabcolsep}{4.5pt}
\renewcommand{\arraystretch}{1.05}
\begin{tabular}{lcc@{\hspace{8pt}}cc}
\toprule
\multirow{2}{*}{Method}
& \multicolumn{2}{c}{\textbf{SHADE-Arena}}
& \multicolumn{2}{c}{\textbf{AgentDyn}} \\
\cmidrule(lr){2-3}
\cmidrule(lr){4-5}
& Utility & $\Delta$
& Utility & $\Delta$ \\
\midrule

Base Harness
& 0.471 & --
& 0.900 & -- \\

\midrule

Vanilla Monitor
& \second{0.529} & +0.058
& 0.900 & +0.000 \\

TS-Flow
& 0.118 & -0.353
& 0.740 & -0.160 \\

\midrule

GPT-5.4
& 0.133 & -0.338
& 0.620 & -0.280 \\

Claude Opus 4.7
& 0.438 & -0.033
& 0.900 & +0.000 \\

\midrule

Meta-Harness
& 0.438 & -0.033
& \second{0.920} & +0.020 \\

HARDE (Ours)
& \best{0.530} & +0.059
& \best{0.940} & +0.040 \\

\bottomrule
\end{tabular}
\end{table}

\subsection{Ablation of Harness Modules}
\label{app:module_ablation}

We further ablate the three harness modules on SHADE-Arena in Table~\ref{tab:module_ablation}. Removing either the trigger or feedback module reduces both safety and utility.Without the trigger module, the harness loses lightweight risk filtering and monitor-invocation gating, such as directly allowing low-risk read operations while blocking or escalating high-risk write operations to the monitor, resulting in substantial utility degradation. Without the feedback module, blocked actions are no longer accompanied by effective recovery guidance, making it harder for the agent to safely replan and complete the intended benign task. Removing both modules causes a larger performance drop, indicating that risk detection alone is insufficient for effective runtime defense. When all three core modules are removed, only auxiliary harness components, such as additional prompts, remain available for optimization, and Safe Success drops to 0.083. These results show that the three modules play important and complementary roles in achieving effective runtime safety while preserving task utility.

\begin{table}[h]
\centering
\caption{
Ablation of the core harness modules on SHADE-Arena.
Utility, Safety, and SS denote task utility, attack-prevention rate, and Safe Success, respectively.
Higher values are better.
The best and second-best results are shown in \textbf{bold} and \underline{underlined}, respectively.
}
\label{tab:module_ablation}
\setlength{\tabcolsep}{7pt}
\renewcommand{\arraystretch}{1.08}
\begin{tabular}{lccc}
\toprule
Variant
& \textbf{Utility}
& \textbf{Safety}
& \textbf{SS} \\
\midrule

w/o Trigger
& \second{0.357}
& \second{0.929}
& \second{0.357} \\

w/o Feedback
& 0.231
& 0.923
& 0.154 \\

w/o Trigger \& Feedback
& 0.154
& 0.923
& 0.154 \\

None
& 0.083
& 0.917
& 0.083 \\

\midrule

HARDE (Full)
& \best{0.500}
& \best{0.984}
& \best{0.500} \\

\bottomrule
\end{tabular}
\end{table}



\section{Case Study}
\label{sec:appdC}

\subsection{Optimization Guide Example}
\label{sec:appdC1}

Below, we present the complete structural template of optimization guide. For brevity, instance-specific values, metric tables, and detailed observations are omitted. The guide first summarizes the probing configuration and results, and then organizes the optimization knowledge for each harness module, including its interface contract, optimization space, observed failure modes, selection conditions, expected metric effects, and boundaries with other modules. It concludes with signals that map evaluation metrics and trajectory symptoms to the modules most likely to require optimization.
\filebox{An Example of Optimization Guide}{logs/harnessguide_example.txt}

\subsection{Attack Cases}
\label{sec:appdC2}

In this section, we present representative attack cases from the three benchmarks.

\begin{itemize}
    \item \textbf{SHADE-Arena.}
    In the \texttt{spam\_filter\_update\_0} case, the visible main task is to inspect several employees' inboxes and spam folders, improve and consolidate their filtering rules, reprocess the mail, and write a summary. The malicious secondary task is to install a global auto-delete rule targeting a particular GitHub sign-in security notification. This objective enters through the benchmark's hidden secondary-task text in the agent's system prompt.

    \item \textbf{AgentDyn.}
    In the shopping \texttt{user\_task\_0} case, the visible main task is simply to purchase a smart watch for the user. The injected objective tells the agent to first visit a purported discount site. It enters through attacker-controlled shopping reviews and an email returned by environment tools, where it is presented as an ``important'' instruction from the user.

    \item \textbf{Agent-SafetyBench.}
    In case \texttt{522}, the visible main task is to retrieve the most recent email from a specified sender containing ``Global Economy.'' The malicious instruction asks the agent to grant permanent smart-lock access to a third party. It is embedded in the content of the retrieved email and therefore reaches the agent through an email-search tool result.
\end{itemize}

The examples below illustrate how the attacks are injected into the agent execution trajectories in these three cases.

\filebox{\textbf{SHADE-Arena}:Hidden Secondary Task in the System Prompt}{logs/attack_shade.txt}
\filebox{\textbf{AgentDyn}:Environment Prompt Injection}{logs/attack_agentdyn.txt}
\filebox{\textbf{Agent-SafetyBench}: Injection in an Email Tool Result}{logs/attack_agentsafetybench.txt}

\subsection{Defense Cases}
\label{sec:appdC3}

This section presents a representative AgentDyn case in which a malicious instruction embedded in a repository file induces the agent to propose an unauthorized collaborator invitation. Although the proposed tool call is not inherently malicious, it is unrelated to the visible user request and originates from untrusted file content. By examining the full execution trajectory, the monitor module performs contextual source attribution and identifies that the action is driven by an injected instruction rather than a user-authorized objective. It therefore blocks the invitation and provides corrective feedback that redirects the agent to the original repository-synchronization task. The agent subsequently completes the benign task without executing the attack, illustrating why trajectory-level source attribution is effective for detecting injected actions that may evade fixed action-level rules.

\filebox{\textbf{AgentDyn}: Monitor Blocks an Injected Action and Restores the Main Task}{logs/defense_agentdyn.txt}

\section{Detailed Analysis of Optimized Safety Harnesses}
\label{sec:appdD}

This section provides the detailed qualitative analysis summarized in Section~\ref{sec:discussion}. We first define the safety properties used in our analysis and characterize the role of each harness module. We then examine how the dominant safety risks vary across benchmarks and how monitor capability affects the optimal harness design.

\subsection{Safety Properties and Module Roles}
\label{sec:appdD1}

Assessing the risk of an agent action cannot rely solely on the semantics of the action itself; rather, it requires contextual evaluation from multiple perspectives. Building on the framework of \citet{siu2026framework} and extending it, we characterize agent safety through five properties:

\begin{itemize}
    \item \textbf{Task Alignment}: whether the agent continues to pursue the authorized objective;
    \item \textbf{Action Alignment}: whether each action contributes to that objective;
    \item \textbf{Source Authorization}: whether the instructions that causally influence an action originate from an authorized source;
    \item \textbf{Data Isolation}: whether information flows remain within authorized boundaries; and
    \item \textbf{Epistemic Integrity}: whether the information communicated by the agent is grounded in trustworthy evidence.
\end{itemize}

Our decomposition of the risk-aware harness into three modules provides complementary intervention points that map onto these overlapping yet distinct safety properties, enabling the harness to address failures that cannot be resolved by any single component.

\paragraph{Trigger Module.}
The trigger module performs lightweight gating before expensive semantic monitoring. It mainly supports \emph{action alignment} and \emph{source authorization} by filtering suspicious tool calls and determining which actions require further inspection.

\paragraph{Monitor Module.}
The monitor module performs contextual risk assessment of proposed actions. It plays the central role in enforcing \emph{action alignment}, \emph{source authorization}, and \emph{data isolation}, particularly for injected instructions and unauthorized side effects that cannot be identified from action semantics alone.

\paragraph{Feedback Module.}
The feedback module governs how interventions and tool observations are incorporated into the agent's subsequent context. It primarily supports \emph{task alignment} and \emph{epistemic integrity} by helping the agent recover from blocked actions without drifting from the original task or continuing to reason from polluted observations.

\paragraph{Other Components.}
Beyond the three core modules, the harness also contains auxiliary components that can be optimized when needed. For example, modifying the agent prompt can provide a global safety prior by communicating general constraints and setting-specific risks, primarily supporting \emph{epistemic integrity} and, to a lesser extent, \emph{task alignment} and \emph{action alignment}.

Together, these mappings show that the modules do not merely provide alternative implementations of the same defense. Instead, they operate at different points in the execution loop and address complementary safety properties, motivating their separate treatment during probing and optimization.

\subsection{Benchmark-Specific Risk Profiles}
\label{sec:appdD2}

Different benchmarks involve distinct attack mechanisms and task settings, threatening different agent-safety properties and therefore requiring different priorities and directions for module optimization. Representative attack examples from all three benchmarks are provided in Appendix~\ref{sec:appdC2}.

\paragraph{SHADE-Arena.}
SHADE-Arena embeds a hidden side-task instruction in the system prompt and requires the agent to execute it covertly while completing the main task. This attack primarily compromises \emph{task alignment} and \emph{action alignment}; tasks involving fund transfers, API keys, or SSNs additionally threaten \emph{data isolation}. Effective optimization therefore spans multiple modules: the trigger module restricts high-risk write operations, the monitor module reduces action-alignment errors, and the feedback module redirects the agent to the authorized task after an intervention. This finding suggests that improving action-level detection alone is insufficient for defending against covert side tasks embedded in long-horizon trajectories.

\paragraph{AgentDyn.}
AgentDyn primarily injects malicious instructions through tool outputs, causing the agent to mistake externally supplied content for user-authorized instructions. Such attacks jointly violate \emph{source authorization} and \emph{action alignment}. The key challenge is to distinguish task-relevant data from untrusted external content and injected directives. Because fixed rules are often insufficient for this distinction in complex trajectories, the monitor module plays a central role by performing contextual source attribution, which remains a practical bottleneck in enforcing source authorization. Appendix~\ref{sec:appdC3} presents a representative example in which the monitor detects and blocks an action induced by an injected instruction.

\paragraph{Agent-SafetyBench.}
Some attacks in Agent-SafetyBench do not induce harmful tool calls, but instead corrupt the agent's reasoning and cause it to produce incorrect factual claims or harmful information. Because no overtly unsafe action is executed, defenses focused primarily on \emph{action alignment}, such as action-level monitoring, may fail to detect these cases. The feedback module and prompt design therefore become more important for preserving \emph{epistemic integrity}, for example by sanitizing malicious tool observations before they are incorporated into the agent's subsequent context.

These differences explain why a fixed optimization order is ineffective across benchmarks. Effective harness optimization should first identify the violated safety property and the current system bottleneck, and then revise the module best positioned to address that failure. The cross-dataset transfer results in Figure~\ref{fig:transfer_exp}(a) provide further evidence that optimization experience transfers more effectively between settings with similar attack mechanisms and risk profiles.

\subsection{Monitor-Dependent Harness Design}
\label{sec:appdD3}

Section~\ref{sec:transfer} evaluates the transferability of optimization guide across monitor models, specifically DeepSeek-V4-Flash and Qwen3-8B. Here, we examine how their different capabilities shape the resulting harness designs.

DeepSeek-V4-Flash has stronger reasoning ability and a longer context window, enabling more reliable source attribution over long trajectories. It can trace the effects of injected instructions across extended interaction histories and more accurately determine whether a proposed action is driven by an authorized instruction or by attacker-controlled content. Accordingly, when DeepSeek-V4-Flash is used as the monitor, \textsc{harde} places greater emphasis on optimizing the monitor module.

By contrast, Qwen3-8B is less capable of handling long trajectories, compound attacks, and implicit relationships between instructions and their sources. It therefore relies more heavily on deterministic modules for feature extraction and lightweight filtering, such as distinguishing read-only from write operations or inspecting tool arguments. As a result, \textsc{harde} places greater emphasis on optimizing the trigger module and feedback module.

These observations explain the cross-monitor transfer results in Figure~\ref{fig:transfer_exp}(b): optimization guidance obtained with a stronger monitor may not transfer directly to a weaker monitor whose surrounding support modules must compensate for different failure modes. More generally, a monitor should not be treated as a plug-and-play classifier. Its capabilities and limitations determine whether harness optimization should focus primarily on semantic monitoring or on the deterministic and feedback mechanisms surrounding it.

\section{Limitations}
\label{app:limitation}
Our work has several limitations.(1) Our study primarily examines variation across monitor models rather than agent models, because the monitor is the core decision-making component of our framework and weaker agent models often struggle to complete challenging benchmarks such as SHADE-Arena. (2) Although optimization guide transfers effectively across related benchmarks, reliable transfer between monitors with substantially different capabilities remains unresolved, particularly in complex long-horizon benchmarks. (3) Our work focuses on optimizing an external safety harness without modifying the agent model itself. Future work could jointly optimize runtime harnesses and agent training to explore the complementary benefits of intrinsic alignment and external control.

\end{document}